\documentclass[aps,amsfonts,amsmath,prd,preprint,nofootinbib]{revtex4-2}
\usepackage{epsfig,bbm,cancel}
\usepackage{slashed,xcolor}
\usepackage[normalem]{ulem} 
\usepackage{amssymb}
\usepackage{amsmath}
\usepackage{amsthm}
\usepackage{mathtools}

\begin{document}

\title{Electric-dual BPS Vortices in The Generalized Self-dual Maxwell-Chern-Simons-Higgs Model}

\author{Laurenzius Yudha Prasetya Tama$^1$}
\email{yudhaprasetyatama@gmail.com}
\author{Bobby Eka Gunara$^1$}
\email{bobby@fi.itb.ac.id (Corresponding author)}
\author{Ardian Nata Atmaja$^2$}
\email{ardi002@brin.go.id}

\affiliation{$^1$Theoretical High Energy Physics  Research Division, Institut Teknologi Bandung,
Jl. Ganesha 10 Bandung 40132, Indonesia.}
\affiliation{$^2$ Research Center for Quantum Physics, National Research and Innovation Agency(BRIN),
Kawasan PUSPIPTEK Serpong, Tangerang 15314, Indonesia.}

\begin{abstract}
In this paper we show how to derive the Bogomolny's equations of the generalized self-dual Maxwell-Chern-Simons-Higgs model presented in  \cite{Bazeia:2012ux} by using the BPS Lagrangian method with a particular choice of the BPS Lagrangian density. We also show that the identification, potential terms, and Gauss's law constraint can be derived rigorously under the BPS Lagrangian method. In this method, we find that the potential terms are the most general form that could have the BPS vortex solutions. The Gauss's law constraint turns out to be the Euler-Lagrange equations of the BPS Lagrangian density. We also find another BPS vortex solutions by taking other identification between the neutral scalar field and the electric scalar potential field, $N=\pm A_0$, which is different by a relative sign to the identification in \cite{Bazeia:2012ux}, $N=\mp A_0$. Under this identification, $N=\pm A_0$, we obtain a slightly different potential terms and Bogomolny's equations compared to the ones in \cite{Bazeia:2012ux}. Furthermore we compute the solutions numerically, with the same configurations as in \cite{Bazeia:2012ux}, and find that only the resulting electric field plots differ by sign relative to the results in \cite{Bazeia:2012ux}. Therefore we conclude that these BPS vortices are electric-dual BPS vortices of the ones computed in \cite{Bazeia:2012ux}.
\end{abstract}

\maketitle
\thispagestyle{empty}
\setcounter{page}{1}
\tableofcontents

\newpage

\section{Introduction}

In the classical picture of  some certain nonlinear field theories we may have  a solution called soliton which differs from the usual elementary particles that are obtained from the quantization of the wave-like excitation of the fields. This object has topological structure and finite energy that is localized in space, see for example,  \cite{manton_sutcliffe_2004}. There are  several kind of solitons: kink, vortex, domain walls, and monopole. Some of them are endowed with a potential that possess a nontrivial vacuum structure,  see for example,  \cite{weinberg_2012}. 

Particularly, our interest turns to consider solitons with co-dimension two or known as vortices in the model with $U(1)$ symmetry. These vortices can be  classified  into two categories, namely, global and gauged vortices that corresponds to the global and local $U(1)$ transformations, respectively. The energy density for global vortices diverge at large $r$. However, the existence of these solutions are still physically relevant in three dimensions such as the cosmic strings \cite{Kibble_1976}. In gauged vortices, the $U(1)$ gauge field is added to the Lagrangian so that it remains invariant. With the addition of gauge field, the divergence that occurs in the energy density vanishes so that the total energy to this solution is proportional to the magnetic flux. Due to its finite energy solution, gauged vortices have interesting application in the superconductors of the second group \cite{Abrikosov:1956sx,NIELSEN197345}.

In the canonical model of gauged vortices, the kinetics of the gauge field is described by the Maxwell term. Nevertheless, there are another possible kinetic term to the gauge field that remains gauge invariant, that is, the Chern-Simons term. In general, this term could exist in any odd-dimensional spacetime. Particularly, in $2+1$-dimensional spacetime it has a quadratic form in the gauge field \cite{10.1007/3-540-46637-1_3}. In the canonical model with only the Maxwell term, there are well-known stable vortices with some particular choice of coupling constants and potential, which turned out to be solutions of first-order differential equations known as the Bogomolny's equations~\cite{NIELSEN197345,osti_7309001}. On the other hand, the canonical model with only the Chern-Simons term also found to have stable vortices~\cite{Paul:1986ix}. For the canonical model with both Maxwell and Chern-Simons terms, the existance of additional neutral scalar filed is necessary in order to have stable vortices~\cite{LEE199079}. The first-order differential formulation in this study leads into a conclusion that there is an identification between the scalar gauge potential and the neutral scalar field. Moreover, there is a recent study that generalizes this idea by introducing coupling functions of the scalar fields to the Maxwell term and all kinetic terms of the scalar fields~\cite{Bazeia:2012ux}. In this generalization, the coupling functions between Maxwell term and the kinetic term of the neutral scalar field are identical. This allows identification between the neutral scalar field and the scalar gauge potential field to be hold which later give us the Bogomolny's equations and their solutions, or known as BPS vortices.

In this paper, we investigate BPS vortices in the generalized self-dual Maxwell-Chern-Simons-Higgs model in \cite{Bazeia:2012ux} using the BPS Lagrangian method developed in \cite{NATAATMAJA2017351}. The BPS Langrangian method is used to find a set of first-order differential equations, called Bogomolny's equations, that solve the second-order differential equations of the equations of motion in the aforementioned generalized self-dual Maxwell-Chern-Simons-Higgs model. We would like to answer a question if the BPS Lagrangian method could reproduce rigorously all the results in \cite{Bazeia:2012ux} and furthermore if there are other possible potential terms that could lead to different Bogomolny's equations and BPS vortices.

This paper consist of four sections including this one. In Sec. \ref{sec:gMCSH}, the radially symmetric ansatz is applied to the Lagrangian density of this model. In Sec. \ref{sec:LBPS-method}, the analysis to obtain the BPS equations is presented. The solution to the obtained BPS equations is done by numerical analysis.  We conclude this study in Sec. \ref{sec:conclusion}, where we give our final comment to the obtained solution and discuss the possibility for further research.


\section{Generalized Self-dual Maxwell-Chern-Simons-Higgs Model}\label{sec:gMCSH}
The model that has been presented in \cite{Bazeia:2012ux} is described by the Lagrangian density of the form
\begin{equation}
    \begin{split}
        \mathcal{L} &= -\frac{h(|\phi|,N)}{4}F_{\mu\nu}F^{\mu\nu} - \frac{\kappa}{4}\epsilon^{\mu\rho\sigma}A_{\mu}F_{\rho\sigma} + w(|\phi|)|D_{\mu}\phi|^{2} + \frac{h(|\phi|,N)}{2}\partial_{\mu}N\partial^{\mu}N - V(|\phi|,N) ,
    \end{split}
\end{equation}
where $h(|\phi|,N)$ and $w(|\phi|)$ are dimensionless positive functions of the scalar fields, $F_{\mu\nu} = \partial_{\mu}A_{\nu} - \partial_{\nu}A_{\mu}$ is the usual electromagnetic field strength tensor, $D_{\mu}\phi = \partial_{\mu}\phi + ieA_{\mu}\phi$ is the covariant derivative of the Higgs field, and neutral scalar field $N$ was added to stabilize the self-dual solutions~\cite{Bazeia:2012ux}. In this paper, we first set the potential $V(|\phi|,N)$ to be arbitrary.

The dynamics of the fields in this model are described by the Euler-Lagrange equations. The equations of motion for the gauge fieds are given by
\begin{equation}\label{EL-Gauss}
    \partial_{\mu}\big(h F^{\mu\nu}\big) + J^{\nu} = \kappa F^{\nu} ~ ,
\end{equation}
where $J^{\nu} = iew\big((D^{\nu}\phi)^*\phi - \phi^*D^{\nu}\phi\big)$ is the electromagnetic current and $F^{\nu} = (1/2)\epsilon^{\nu\rho\sigma}F_{\rho\sigma}$ is the electromagnetic dual vector. From this equation, one can show that the temporal gauge, $A_{0} = 0$, cannot be used since it will lead to the trivial solution. The remaining equations of motion for the scalar fields, $\phi$ and $N$ respectively, are
\begin{align}
            &D_{\mu}\big(w(|\phi|)D^{\mu}\phi\big)+ \frac{1}{2}\partial_{\phi}h\left( F_{\mu}F^{\mu} -\partial_{\mu}N\partial^{\mu}N \right)- \partial_{\phi}w|D_{\mu}\phi|^{2}+ \partial_{\phi}V = 0\label{EL-Higgs}\\
    &\partial_{\mu}\big(h\partial^{\mu}N\big) + \frac{1}{2}\partial_{N}h\big(F_{\mu}F^{\mu} - \partial_{\mu}N\partial^{\mu}N\big) + \partial_{N}V = 0. \label{EL-NScalar}
\end{align}
In order to solve these equations of motion we impose the usual radially symmetric ansatz
\begin{equation}\label{ansatz-1}
    \phi(r,\theta) = g(r)e^{in\theta} ~ , \quad N\equiv N(r) ~ , \quad A_{0} \equiv A_{0}(r) ~ ,
\end{equation}
\begin{equation}\label{ansatz-2}
    \textbf{A}(r,\theta) = -\frac{\hat{\theta}}{r}\left(a(r)-n\right) ~ ,
\end{equation}
where $n=\pm1,\pm2,\pm3,\ldots$ is the winding number.
Hereafter, we will use the metric $\eta^{\mu\nu} = \text{diag}(+,-,-)$ and a unit such that $e=v=\kappa = 1$, where $v$ is the vacuum expectation value of the Higgs scalar field. Under the ansatz~\ref{ansatz-1}, the Lagrangian density of this model can be effectively written as
\begin{equation}
    \mathcal{L} = \frac{h}{2}{A'}_{0}^{2} - \frac{h}{2}\left(\frac{a'}{r}\right)^{2} - A_{0}\frac{a'}{r} - w\left({g'}^{2} + \frac{{a^2}{g^2}}{r^2}\right)+ {A_0^2}{g^2}w - \frac{h}{2}{N'}^{2} - V(g,N),
\end{equation}
where $h\equiv h(g,N)$, $w\equiv w(g)$, and $'\equiv {d\over dr}$.

\section{BPS Lagrangian Method}\label{sec:LBPS-method}

It has been shown in \cite{NATAATMAJA2017351} that the BPS Lagrangian method was able to rederive the Bogomolny's equations  of the Maxwell-Higgs and Born-Infeld-Higgs models. This method uses the fact that the (effective) Lagrangian density of a system must contains at most up to quadratic of first-derivative of the effective fields. Under this requirement the effective Lagrangian density of a model with $N$-scalar fields can be rewritten into the following form
\begin{equation}\label{general-lagrangian}
    \mathcal{L} = \sum_{i = 1}^{N}\left(\partial_{\mu}{\phi^i} - f_\mu^{i}\left(\phi^j,\partial_{\nu}\phi^{j};x^\nu\right)\right)^{2} + \mathcal{L}_{BPS} ~ ,
\end{equation}
where in general $f_\mu^i$ is a function of fields $\phi^j$'s and their first-derivative $\partial_\nu\phi^j$'s, and may also explicit coordinates $x^\nu$, but not of $\partial_\mu\phi^i$, with $j=1,\ldots,N$. Here we shall call $\mathcal{L}_{BPS}$ as BPS Lagrangian density which in general is a function is a function of fields $\phi^j$'s and their first-derivative $\partial_\nu\phi^j$'s. For most of the known cases, $\mathcal{L}_{BPS}$ contains at most first-derivative of the effective fields up to the power one. The Bogomolny's equations are obtained in the limit where $\mathcal{L}-\mathcal{L}_{BPS}=0$, or also known as the BPS limit condition, which, from \eqref{general-lagrangian}, are given by
\begin{equation}
 \partial_{\mu}\phi^i = f_\mu^{i}\left(\phi^j,\partial_{\nu}\phi^{j},x^\nu\right).
\end{equation}

The BPS Lagrangian density plays an important role in the BPS Lagrangian method. Its Euler-Lagrange equations, which are called constraint equations, must be considered in finding solitonic solutions, in addition to the Bogomolny's equations. Depending on the choice of terms in the BPS Lagrangian density, its Euler-Lagrange equations could be all trivial in the BPS limit. In this case the BPS Lagrangian density posses boundary terms such that it does not contribute to the equations of motion of the model. Type of boundary terms that could be included in the BPS Lagrangian density can be found in~\cite{Adam:2016ipc}.
In general, the BPS Lagrangian density containing both boundary and non-boundary terms can be written as follows
\begin{align}
\begin{split}
        \mathcal{L}_{BPS} =& \frac{1}{\sqrt{\det(g_{mn})}}\left(D_{\mu}F^{\mu}(\phi^a,x^\nu) + \sum_{j = 2}^{\min(N,d)}F^{\mu_{1}...\mu_{j}}_{{a_1}...{a_j}}(\phi^a)~\phi^{a_{1}}_{,\mu_{1}}...\phi^{a_{j}}_{,\mu_{j}}\right)\\
        &+ \sum_{p,q = 1}^{N}X_{p q}^{\mu_1\mu_2}(\phi^a,x^\nu){\partial_{\mu_1}\phi^{p}}{\partial_{\mu_2}\phi^{q}} ~ ,
        \end{split}
\end{align}
where $F^{\mu_{1}...\mu_{j}}_{{a_1}...{a_j}}$ is totally anti-symmetric tensor in all indices, $D_{\mu}$ is total derivative, and the notation $_{,\mu_1} \equiv \partial_{\mu_{1}}$. The function $X_{pq}^{\mu_1\mu_2}$ is not totally anti-symmetric tensor which correspond to the non-boundary terms of the BPS Lagrangian density. Previous study has been made regarding this non-boundary term of the BPS Lagrangian in \cite{Atmaja:2018ddi} which results in non-zero stress tensor of the generalized Born-Infeld-Higgs model.

\subsection{Bogomolny's equations}
For the purpose of rederiving the Bogomolny's equations of this model as in~\cite{PhysRevD.85.125028}, with regard to the defined ansatze in (\ref{ansatz-1}) and (\ref{ansatz-2}), we find the BPS Lagrangian density of the form
\begin{equation}\label{LBPS}
    \begin{split}
        \mathcal{L}_{BPS} = -X_{0} - \frac{X_1}{r}{g'} - \frac{X_2}{r}{a'} - {X_3}~{N'}^{2} - {X_4}~{A'_0}^{2} ~ ,
    \end{split}
\end{equation}
where the $X$'s are undetermined constraint functions that depend only on the effective fields but do not depend explicitly on the coordinates. We then take the BPS limit of $\mathcal{L} - \mathcal{L}_{BPS}$,
\begin{equation}
    \begin{split}
        &w{g'}^{2} - \frac{X_1}{r}{g'} + \frac{h}{2}\left(\frac{a'}{r}\right)^{2} - ({X_2} - {A_0})\frac{a'}{r}+ \left(\frac{h}{2} - X_{3}\right){N'}^{2} - \left(\frac{h}{2} + X_{4}\right){A'}_{0}^{2}\\& + \frac{{a^2}{g^2}w}{r^2} - {A_0^2}{g^2}w - {X_0} + V = 0,
    \end{split}
\end{equation}
 and obtain the Bogomolny's equations.
The above expression can be seen as a quadratic equation in the first-derivative of the fields. At first, we can consider it as a quadratic equation in $g'(r)$. Using the quadratic formula, we obtain
\begin{align}
    {g'}_{\pm} =& \frac{{X_1} \pm r\sqrt{S_1}}{2wr} ~ ,\label{BEg}\\
    \begin{split}
        {S_1} =& 4 w \bigg(-\frac{A_0 a'}{r}-\frac{a^2 g^2 w}{r^2}-\frac{h \left(a'\right)^2}{2 r^2}+\frac{X_2 a'}{r}+A_0^2 g^2 w+\frac{1}{2} h \left(A_0'\right){}^2+X_4 \left(A_0'\right){}^2 \\&-\frac{1}{2} h \left(N'\right)^2  +X_3 \left(N'\right)^2-V+X_0\bigg)+\frac{X_1^2}{r^2} ~ .
    \end{split}
\end{align}
The Bogomolny's equation for $g(r)$ is obtained when $S_1 = 0$. Next, the equation $S_1 = 0$ can be considered as a quadratic equation in $a'(r)$. Repeating the same calculation, we obtain
\begin{align}
    \bigg(\frac{a'}{r}\bigg)_{\pm} =& \frac{4w({X_2}-{A_0})\pm r\sqrt{S_2}}{4h w} ~ , \label{BEa}\\
    \begin{split}
        {S_2} =&\frac{8 h w }{r^2}\bigg(2w {A'_0}^2\left(2X_4+h\right)+2w N'^2\left(2X_3-h\right)\bigg)+\frac{8 h w}{r^4}\left(X_1^2-4 a^2 g^2 w^2\right)\\&+\frac{32 h w^2 }{r^2} \left(X_0-V+\frac{({X_2}-{A_0})^2}{2 h}+A_0^2 g^2 w\right)  ~ .
    \end{split}\label{S2}
\end{align}
Here we can see that there are no linear terms of $N'$ and $A'_0$ in $S_2$. Proceeding further with the same calculation as in the preceding, by setting $S_2=0$, leads to Bogomolny's equations $N' = 0$ and $A'_0 = 0$, along with $g'=X_1/2wr$ and $a'=r({X_2}-{A_0})/h$, and the sum of remaining terms in the right hand side of \eqref{S2} is equal to zero, which can be solved order by order in explicit power of $r$ to obtain
\begin{align}
    \label{X0}&X_0 = V - \frac{({X_2}-{A_0})^2}{2 h} - {A_0^2}{g^2}w ~ ,\\
    \label{X1}&X_1 = \pm 2agw ~ .
\end{align}

There are still three remaining unknown functions in the BPS Lagrangian density \eqref{LBPS}, namely $X_2, X_3$ and $X_4$. On the other hand, the BPS Lagrangian density \eqref{LBPS} implies the constraint equations that must be considered in finding the BPS vortex solutions. Later on we will see these constraint equations can also be used to fix the three unknown functions. Below we will analyse the Euler Lagrange equations of the BPS Lagrangian \eqref{LBPS}.

\subsection{Constraint equations}

Taking the variation on the BPS Lagrangian density for each of the effective fields gives the constraint equations for effective field $g(r)$,
\begin{equation}\label{EL-g}
    \frac{d{X_1}}{d r} = r\frac{\partial X_0}{\partial g} + \frac{\partial X_1}{\partial g}{g'} + \frac{\partial X_2}{\partial g}{a'} + r\frac{\partial X_3}{\partial g}{N'}^{2} + r\frac{\partial X_4}{\partial g}{A'_0}^{2} ~ ,
\end{equation}
for effective field $a(r)$,
\begin{equation}\label{EL-a}
    \frac{d{X_2}}{dr} = r\frac{\partial X_0}{\partial a} + \frac{\partial X_1}{\partial a}{g'} + \frac{\partial X_2}{\partial a}{a'} + r\frac{\partial X_3}{\partial a}{N'}^{2} + r\frac{\partial X_4}{\partial a}{A'_0}^{2} ~ ,
\end{equation}
for effective field $N(r)$,
\begin{equation}\label{EL-N}
    \begin{split}
        \frac{d}{dr}\left(2r{X_3}{N'}\right) = r\frac{\partial X_0}{\partial N} + \frac{\partial X_1}{\partial N}{g'} + \frac{\partial X_2}{\partial N}{a'} \\+ r\frac{\partial X_3}{\partial N}{N'}^{2} + r\frac{\partial X_4}{\partial N}{A'_0}^{2} ~ ,
    \end{split}
\end{equation}
and for effective field ${A_0}(r)$,
\begin{equation}\label{EL-0}
    \begin{split}
        \frac{d}{dr}\left(2r{X_4}{A'_0}\right) = r\frac{\partial X_0}{\partial A_0} + \frac{\partial X_1}{\partial A_0}{g'} + \frac{\partial X_2}{\partial A_0}{a'}\\ + r\frac{\partial X_3}{\partial A_0}{N'}^{2} + r\frac{\partial X_4}{\partial A_0}{A'_0}^{2} ~ .
    \end{split}
\end{equation}

Next, we substitute the $X_0$, $X_1$, and the Bogomolny's equations above into these contraint equations. The equation (\ref{EL-a}) then can be rewritten as
\begin{equation}\label{X2}
    \frac{d X_2}{d r} = \pm 2 g w\frac{d g}{d r} ~ ,
\end{equation}
which implies $X_2 = \pm G(g)$, with
\begin{equation}
    \frac{d G}{d g} = 2 g w ~ .
\end{equation}
Futher substituting this $X_2$ into the remaining constraint equations, the equations \eqref{EL-g},\eqref{EL-N}, and \eqref{EL-0} yield ${\frac{\partial X_0}{\partial g}}={\frac{\partial X_0}{\partial N}}={\frac{\partial X_0}{\partial A_0}}=0$. 
Notice that the functions $X_3$ and $X_4$ are still arbitrary which are the undesirable results. The arbitrariness of these functions is due to the trivial Bogomolny's equations: $A_0'=N'=0$. The case where at least one of the Bogomolny's equations is trivial will be disscussed elsewhere~\footnote{Work in progress.}.

In dealing with these problems, we would take $X_3 = -X_4 = \frac{h}{2}$ rather than $N' =A'_{0} = 0$ in order for the first line of right hand side of equation \eqref{S2} to be zero. This also means the effective fields $A_0$ and $N$ do not have their Bogomolny's equations or in other words the equations that determined their solutions must be of the second-order differential equations. In this way we also fix all the unknown functions \footnote{The function $X_2$ is determined by the equation \eqref{EL-a} as in the previous calculation since $X_3 = -X_4 = \frac{h}{2}$ do not alter the resulting equation \eqref{X2}.} from the BPS Langrangian density \eqref{LBPS}. The remaining constraint equations \eqref{EL-g}, \eqref{EL-N}, and \eqref{EL-0} now become
\begin{equation}\label{Persamaan-g}
    {\frac{1}{2}}\frac{\partial h}{\partial g}\left({N'}^{2} - {A'}_{0}^{2}\right) = \frac{\partial}{\partial g}\bigg(\frac{(G\mp A_0)^2}{2h} + {A_0^2}{g^2}w - V\bigg) ~ ,
\end{equation}
\begin{equation}\label{Persamaan-N}
    \frac{1}{r}\dfrac{d}{d r}\left(r h \dfrac{d N}{d r}\right) = \dfrac{\partial V}{\partial N} + \dfrac{(G\mp A_0)^2}{2h^2}\dfrac{\partial h}{\partial N} + {\frac{1}{2}}\dfrac{\partial h}{\partial N}\left({N'}^{2}-{A'}_{0}^{2}\right) ~ ,
\end{equation}
and
\begin{equation}\label{Persamaan-A0}
    \frac{1}{r}\frac{d}{d r}\bigg(r h \frac{d A_0}{d r}\bigg) = \mp \frac{(G\mp A_0)}{h} + 2{A_0}{g^2}w ~ ,
\end{equation} 
respectively.

At this stage we have three constraint equations in which two of them, \eqref{Persamaan-N} and \eqref{Persamaan-A0}, determine solutions for the effective fields $N$ and $A_0$. The constraint equation \eqref{Persamaan-g} is problematic because it may  further constraint the solutions for effective fields $N$ and $A_0$ that could make the system of constraint equations to be overdetermined. It also contradicts with our premise that there are no Bogomolny's equations for the effective fields $N$ and $A_0$. Therefore we must force the constraint equation \eqref{Persamaan-g} to be trivially satisfied or, at least, eliminate all terms containing $A_0'$ and $N'$ in the constraint equation \eqref{Persamaan-g}.
A possible way to tackle this problem, that does not lead to inconsistency, is by identifying $A_0=\mp N$ \footnote{Here we take the signature to be the same as in $X_2$. We may also identify it another way around by taking $N=\mp A_0$, but this will make the analysis on the potential $V$ turns out to be more difficult.}. Substituting this into the equations \eqref{Persamaan-g}, \eqref{Persamaan-N}, and \eqref{Persamaan-A0} gives us 
\begin{equation}\label{Vg}
 \frac{\partial V}{\partial g}= \frac{\partial}{\partial g}\left(\frac{(G+N)^2}{2h} + N^2g^2w\right) ~ ,
\end{equation}
\begin{equation}\label{N1}
 \frac{1}{r}\dfrac{d}{d r}\left(r h \dfrac{d N}{d r}\right) = \dfrac{\partial V}{\partial N} + \dfrac{(G+N)^2}{2h^2}\dfrac{\partial h}{\partial N} ~ ,
\end{equation}
and 
\begin{equation}\label{N2}
 \frac{1}{r}\frac{d}{d r}\bigg(r h \frac{d N}{d r}\bigg) = \frac{(G+N)}{h} + 2N{g^2}w ~,
\end{equation}
respectively.
The equations \eqref{N1} and \eqref{N2} must be identical and so from these equations we get
\begin{equation}\label{VN}
 \frac{\partial V}{\partial N}=\frac{\partial}{\partial N}\left(\frac{(G+N)^2}{2h} +N^2g^2w\right) ~ .
\end{equation}
From the equations \eqref{Vg} and \eqref{VN}, we may conclude
\begin{equation}\label{potential}
 V= \frac{(G+N)^2}{ 2h}+N^2g^2w ~ ,
\end{equation}
and we get the same results as obtained by Bazeia and others in~\cite{PhysRevD.85.125028}. 

There is also another possible way by identifying $A_0=\pm N$. Following the same steps as in the previous identification, one can easily show that the potential is given by
\begin{equation}\label{new-potential}
 V= \frac{(G-N)^2}{2h}+N^2g^2w ~ ,
\end{equation}
with the Bogomolny's equations
\begin{equation}\label{BPSpm-g}
 g'=\pm \frac{ag}{r} ~ ,
\end{equation}
and
\begin{equation}\label{BPSpm-a}
  \frac{a'}{ r}=\pm \frac{(G-N)}{h} ~ ,
\end{equation}
and the constraint equation
\begin{equation}\label{Npm}
 \frac{1}{r}\frac{d}{d r}\bigg(r h \frac{d N}{d r}\bigg) = -\frac{(G-N)}{h} + 2N{g^2}w ~ ,
\end{equation}
which is simply the Gauss's law constraint equation upon substituting $N=\pm A_0$.

\subsection{Numerical solutions}
In this section, we would like to find solutions for all effective fields by solving the equations (\ref{BPSpm-g}), (\ref{BPSpm-a}), and (\ref{Npm}) numerically with the explicit form of $w(g)$ and $h(g,N)$ given in the Ref. \cite{Bazeia:2012ux},
\begin{eqnarray}
 &&w(|\phi|) = b(|\phi|^2 - 1)^{b-1} ~ ,\\
 &&h(|\phi|,N) = \alpha{N^2} + 1 ~ ,
\end{eqnarray}
where $\alpha$ is a real non-negative number and $b$ is a positive odd number. The effective fields must satisfy the following boundary conditions
\begin{eqnarray}
    &&g(0) = 0 ~ ,\qquad a(0) = n ~ ,\qquad N'(0) = 0 ~ ,\nonumber\\
    &&g(\infty) = 1 ~ ,\qquad a(\infty) = 0 ~ ,\qquad N(\infty) = 0 ~ .
\end{eqnarray}
We solve those equations numerically using the perturbed iterative scheme (PIS) that was introduced in Ref. \cite{DEY197715}. Then we compute the BPS energy density, magnetic field, and electric field via
\begin{eqnarray}
&&\rho_{BPS} = \pm\frac{1}{r}\frac{d}{d r}(a G) + \frac{1}{r}\frac{d}{d r}\left(r h N \frac{d N}{d r}\right) ~ ,\\
&&B(r) =F_{12}= -\frac{1}{r}\frac{d a}{d r} ~ ,
\end{eqnarray}
and
\begin{equation}
    E(r) =F_{0r}=-\frac{d A_0}{d r} ~ ,
\end{equation}
respectively.

\begin{figure}[h]
    \centering
    \includegraphics[width=0.6\textwidth]{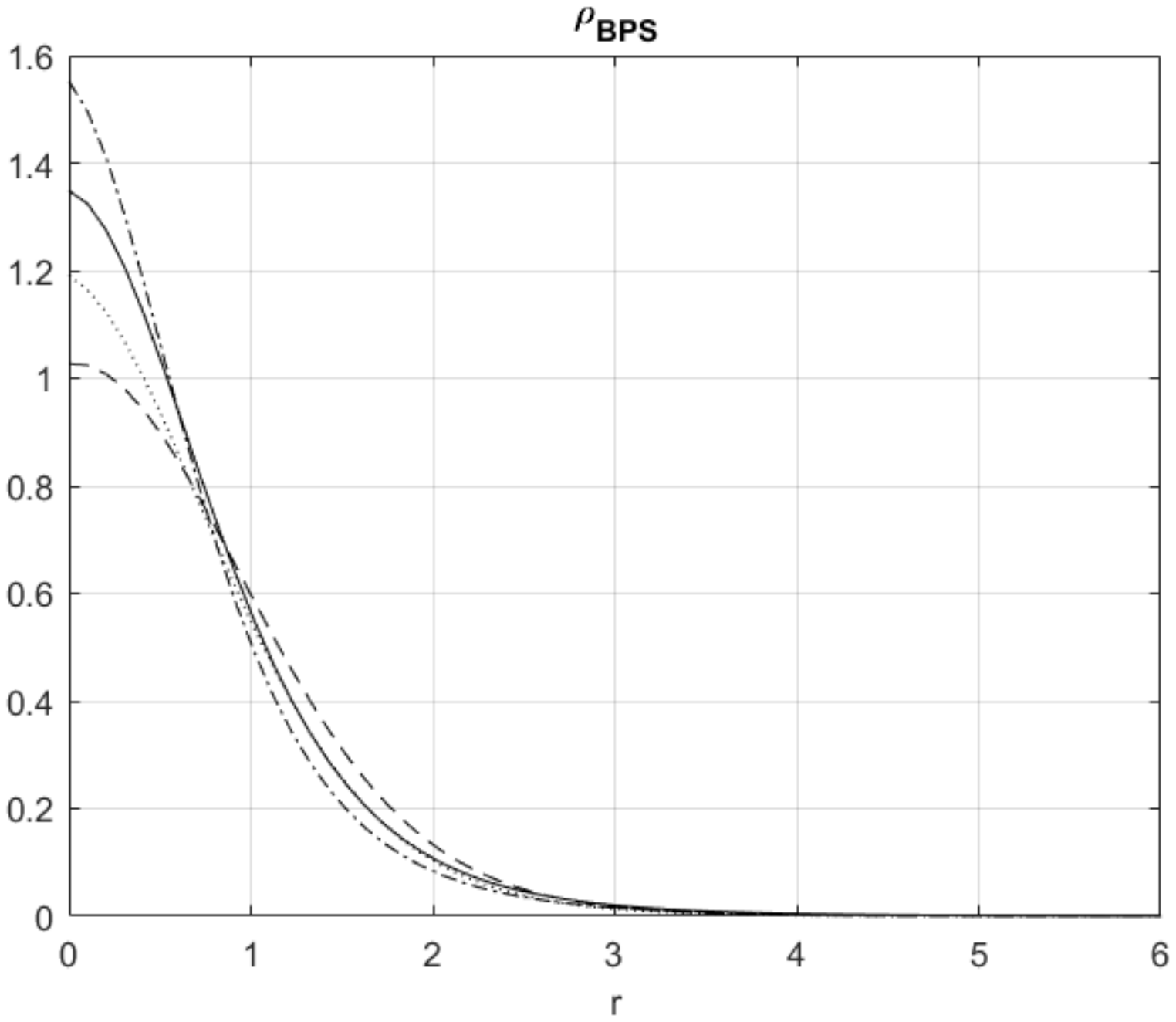}
    \caption{Plots of the BPS energy density, $\rho$(r), with $\alpha = 0$ and $b=1$ (dash-dotted line), $\alpha = 0$ and $b=3$ (dotted line), $\alpha = 5$ and $b=1$ (dashed line), and $\alpha = 5$ and $b = 3$ (solid line).}
    \label{fig:Edensity}
\end{figure}

FIG. \ref{fig:Edensity}. shows the energy density of the $(n=1)$ vortex solutions for various value of $\alpha$ and $b$. We plot the solutions for the values of $\alpha = 0$ and $b=1$ (dash-dotted line), $\alpha = 0$ and $b = 3$ (dotted line), $\alpha = 5$ and $b=1$ (dashed line), and $\alpha = 5$ and $b = 3$ (solid line). From the dotted and the dashed plots, we can see that as we vary the value of $\alpha$ for a fixed $b$, the amplitude is decreasing. The same situation is also happened for the dash-dotted and the solid plots. From the dotted and the dash-dotted plots, as we vary the value of $b$ for fixed value of $\alpha$, it is visible that the amplitude of the energy density function increases. We can also deduce that as the amplitude increases, the characteristic length of the vortex decreases. With the same values of $\alpha$ and $b$ as the ones used in FIG. 3 of Ref. \cite{Bazeia:2012ux}, we can compare both the graphs and observe that they share same characteristics of the energy density. Hence we may conclude that our choice of sign in the identification does not change the total energy.

\begin{figure}[h]
    \centering
    \includegraphics[width=0.6\textwidth]{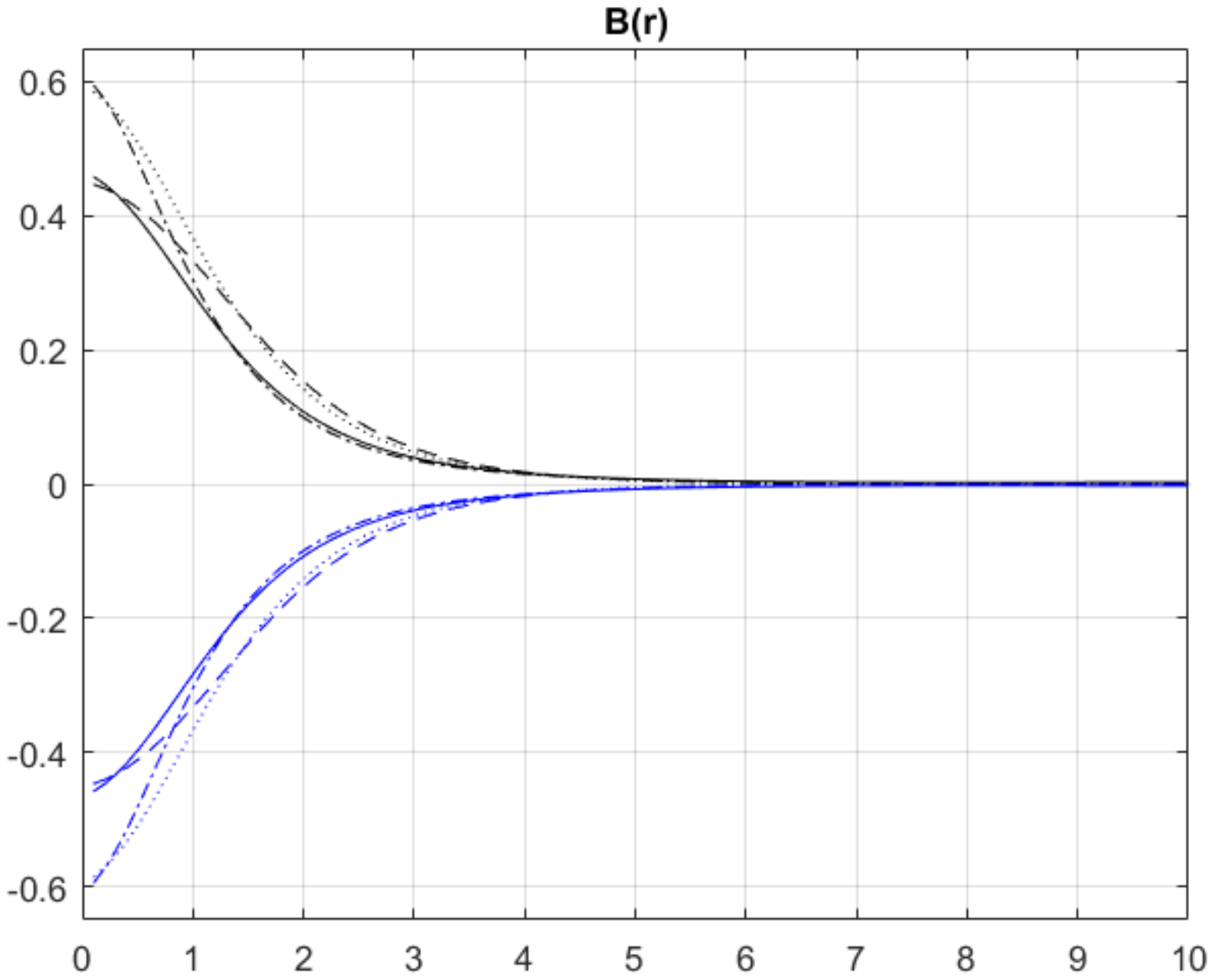}
    \caption{Plots of the magnetic field, $B(r)$, for the $(n = 1)$ vortex solutions (black) and the $(n = -1)$ anti-vortex solutions (blue) with the same values of $\alpha$ and $b$ as in FIG. \ref{fig:Edensity}}
    \label{fig:B(r)}
\end{figure}

In FIG. \ref{fig:B(r)}., we show the plots of magnetic field with the same values of $\alpha$ and $b$ as in FIG. \ref{fig:Edensity}. We calculate both the $(n=1)$ vortex  and $(n=-1)$ anti-vortex solutions which are depicted by the black and blue plots, respectively. From the graph, we can see that as we increase the value of $\alpha$ and fix the value of $b$, the amplitude of the magnetic field is decreasing. Moreover, varying the parameter $b$, while $\alpha$ is fixed, result in the increase of the amplitude and the decrease of the characteristic length. Again here our graph is similar to the one in FIG. 3 of Ref. \cite{Bazeia:2012ux}, and hence our choice of sign in the identification does not change the profile of magnetic field.

\begin{figure}[h]
    \centering
    \includegraphics[width=0.6\textwidth]{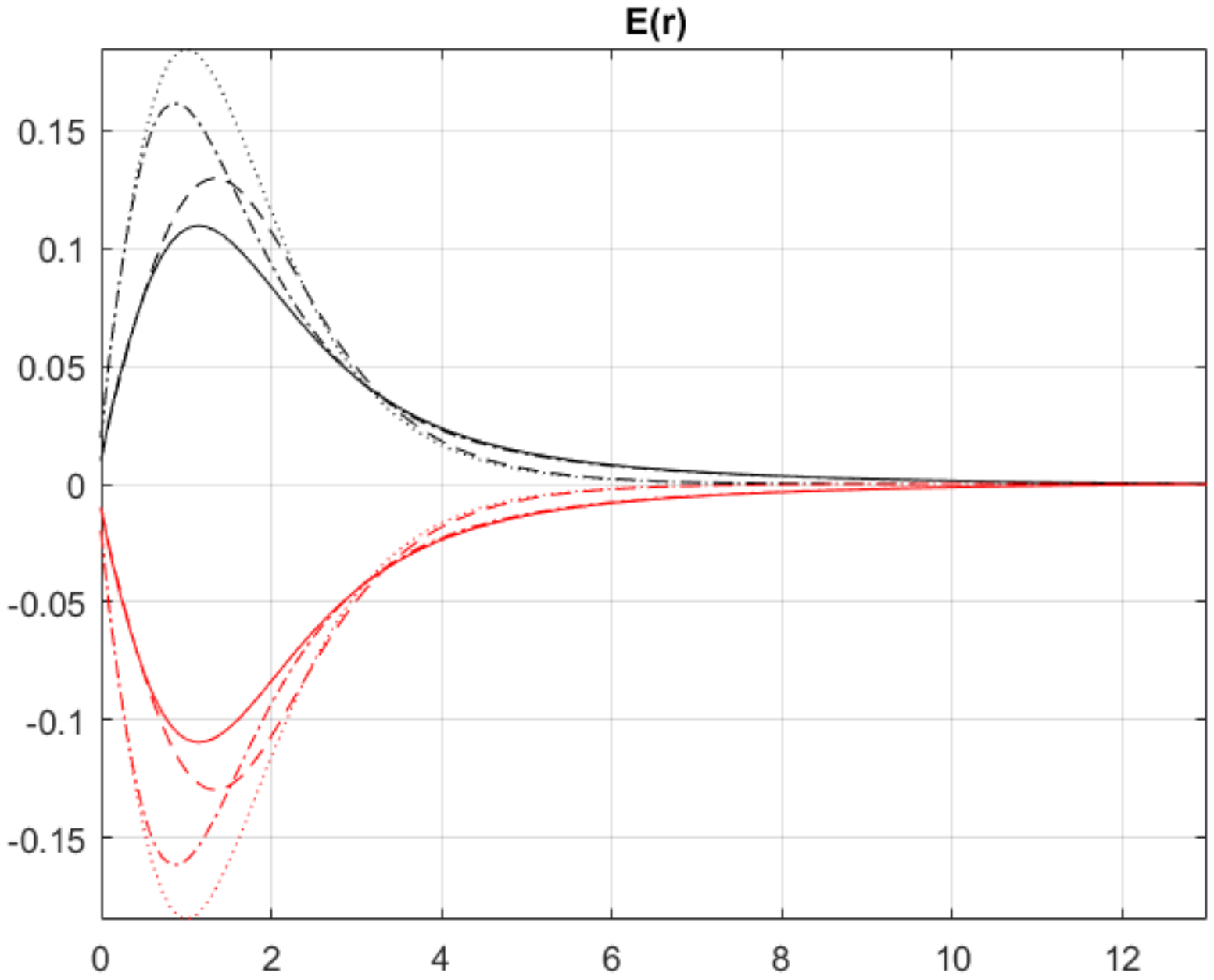}
    \caption{Plots of the electric field, $E(r)$, for $N = \mp {A_0}$ (black) and for $N = \pm A_0$ (red) with the same convention for $\alpha$ and $b$ as in FIG. \ref{fig:Edensity}.}
    \label{fig:E(r)}
\end{figure}

\begin{figure}[h]
    \centering
    \includegraphics[width=0.6\textwidth]{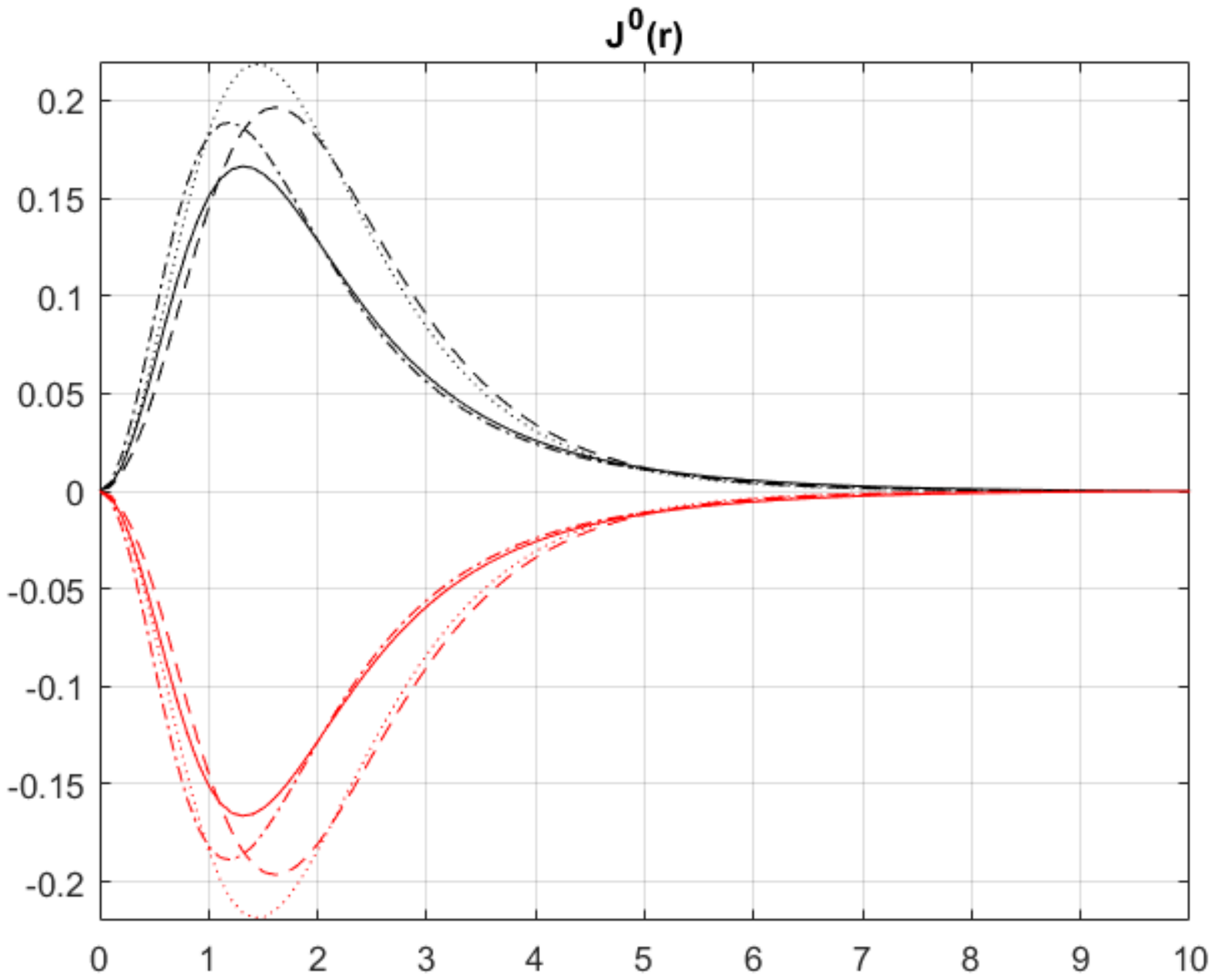}
    \caption{The electric charge density, $J^{0}(r)$, of the vortex solution for $N=\mp A_0$ (black) and for $N = \pm A_0$ (red) with the same convention for $\alpha$ and $b$ as in FIG. \ref{fig:Edensity}}
    \label{fig:J0(r)}
\end{figure}

In FIG. \ref{fig:E(r)}., we show the electric field of the vortex solution $(n=1)$ from the identification $A_0 = \mp N$ (black) and the identification $A_0 = \pm N$ (red) with the convention for the parameters $\alpha$ and $b$ is the same as in FIG. \ref{fig:Edensity}. We observe that each electric field of the vortices between these two identifications, for the same values of $\alpha$ and $b$, are different by sign. The electric field plots in Ref.\cite{Bazeia:2012ux} are related to the electrically positive charged BPS vortex solutions, while our plots, with identification $N = \pm A_0$, are related to the corresponding electrically negative charged BPS vortex solutions. This fact can be observed more clearly from the plots of electric charge density $J^0$ in the FIG. \ref{fig:J0(r)}. Here we can see that the electric charge density is localized from which the electric charge is defined by
\begin{equation}
    Q_e = {2\pi\over h_\infty} \int^\infty_0 r J^{0}(r) ~ dr ~ ,
\end{equation}
where $h_\infty$ is the asymptotic value of $h(|\phi|,N)$, and $J^0$ is the time component of the electromagnetic current.

\section{Conclusion}\label{sec:conclusion}
We have shown how a particular choice of BPS Lagrangian density \eqref{LBPS} is sufficient to reproduce rigorously all the results in Ref. \cite{Bazeia:2012ux} by using the BPS Lagrangian method \cite{NATAATMAJA2017351}. We would like to point out that the potential \eqref{potential} was derived rigorously under the BPS Lagrangian method and it is the most general potential, relative to the BPS Lagrangian density \eqref{LBPS}, that could bear the BPS vortices. Moreover we did not need to impose a priori the Gauss's law constraint equation, such as in Ref. \cite{Bazeia:2012ux}, since it is also derived rigorously as the constraint equation \eqref{Persamaan-A0}, or similarly \eqref{N2}, of the BPS Lagrangian density \eqref{LBPS} in the BPS limit. 


We also found another possible identification with $A_0=\pm N$, which is different by sign, that leads to potential \eqref{new-potential}. We calculated and compared our solutions with the ones in Ref. \cite{Bazeia:2012ux}. As we observed from the FIG. \ref{fig:E(r)}., our solutions describe BPS vortex solutions with negative electric charge which are opposite to the ones obtained in Ref. \cite{Bazeia:2012ux}. One can simply check that the Bogomolny's and constraint equations, along with the potential, can be obtained from the corresponding equations in Ref. \cite{Bazeia:2012ux} by transforming the neutral scalar field $N\to -N$. This also the reason why the BPS vortex solutions have negative electric charge, even if $e=1$, since the electric charge density $J^0\propto N$. Here we may conclude that BPS vortices with identification $A_0=\pm N$ are electrically dual to the BPS vortices with identification $A_0=\mp N$ as found in Ref. \cite{Bazeia:2012ux}.

The BPS Lagrangian density \eqref{LBPS} is not the only possible BPS Lagrangian density that one could consider. The most general BPS Lagrangian density consist of all possible terms would be
\begin{eqnarray}
 \mathcal{L}{BPS}&=&X{0}+ \frac{X_1}{r}{a'} + \frac{X_2}{r}{g'} + \frac{X_3}{r} {N'} + \frac{X_4}{ r} {A'_0}\nonumber\\
 &&+X_5 {a'}^2+X_6 {a'g'}+X_7 {a'N'}+X_8{a'A_0'} \nonumber\\
 &&+X_9 {g'}^2+X_{10}{g'N'}+X_{11}{g'A_0'}\nonumber\\
 &&+X_{12} {N'}^2+X_{13} {N'A_0'}+ X_{14} {A_0'}^2.
\end{eqnarray}
Depending on the choice of terms in the BPS Lagrangian density, the resulting Bogomolny's and constraint equations, as well as the form of potential terms and functions $h$ and $w$, could be different from the ones obtained here. Other possibles BPS Lagrangian density will be studied elsewhere.
\section*{Acknowledgments}

ANA would like to thank the ICTP Associate office for the support, under the ICTP Associate scheme programme (2018-2023), and the hospitaly during his visit to the Abdus Salam ICTP in 2022 where some parts of this article were written. The work in this paper is supported by Kemendikbudristek-ITB WCR 2021 and PDUPT Kemendikbudristek-ITB 2022. The work of BEG is also supported by Riset ITB 2022.  B. E. G. would also acknowledge the support from the ICTP through the Associates Programme (2017-2022). 

\medskip
$\bibliographystyle{utphys}$
\bibliography{references}

\begin{thebibliography}{15}%
\makeatletter
\providecommand \@ifxundefined [1]{%
 \@ifx{#1\undefined}
}%
\providecommand \@ifnum [1]{%
 \ifnum #1\expandafter \@firstoftwo
 \else \expandafter \@secondoftwo
 \fi
}%
\providecommand \@ifx [1]{%
 \ifx #1\expandafter \@firstoftwo
 \else \expandafter \@secondoftwo
 \fi
}%
\providecommand \natexlab [1]{#1}%
\providecommand \enquote  [1]{``#1''}%
\providecommand \bibnamefont  [1]{#1}%
\providecommand \bibfnamefont [1]{#1}%
\providecommand \citenamefont [1]{#1}%
\providecommand \href@noop [0]{\@secondoftwo}%
\providecommand \href [0]{\begingroup \@sanitize@url \@href}%
\providecommand \@href[1]{\@@startlink{#1}\@@href}%
\providecommand \@@href[1]{\endgroup#1\@@endlink}%
\providecommand \@sanitize@url [0]{\catcode `\\12\catcode `\$12\catcode
  `\&12\catcode `\#12\catcode `\^12\catcode `\_12\catcode `\%12\relax}%
\providecommand \@@startlink[1]{}%
\providecommand \@@endlink[0]{}%
\providecommand \url  [0]{\begingroup\@sanitize@url \@url }%
\providecommand \@url [1]{\endgroup\@href {#1}{\urlprefix }}%
\providecommand \urlprefix  [0]{URL }%
\providecommand \Eprint [0]{\href }%
\providecommand \doibase [0]{https://doi.org/}%
\providecommand \selectlanguage [0]{\@gobble}%
\providecommand \bibinfo  [0]{\@secondoftwo}%
\providecommand \bibfield  [0]{\@secondoftwo}%
\providecommand \translation [1]{[#1]}%
\providecommand \BibitemOpen [0]{}%
\providecommand \bibitemStop [0]{}%
\providecommand \bibitemNoStop [0]{.\EOS\space}%
\providecommand \EOS [0]{\spacefactor3000\relax}%
\providecommand \BibitemShut  [1]{\csname bibitem#1\endcsname}%
\let\auto@bib@innerbib\@empty
\bibitem [{\citenamefont {Bazeia}\ \emph
  {et~al.}(2012{\natexlab{a}})\citenamefont {Bazeia}, \citenamefont {Casana},
  \citenamefont {da~Hora},\ and\ \citenamefont {Menezes}}]{Bazeia:2012ux}%
  \BibitemOpen
  \bibfield  {author} {\bibinfo {author} {\bibfnamefont {D.}~\bibnamefont
  {Bazeia}}, \bibinfo {author} {\bibfnamefont {R.}~\bibnamefont {Casana}},
  \bibinfo {author} {\bibfnamefont {E.}~\bibnamefont {da~Hora}},\ and\ \bibinfo
  {author} {\bibfnamefont {R.}~\bibnamefont {Menezes}},\ }\bibfield  {title}
  {\bibinfo {title} {{Generalized self-dual Maxwell-Chern-Simons-Higgs
  model}},\ }\href {https://doi.org/10.1103/PhysRevD.85.125028} {\bibfield
  {journal} {\bibinfo  {journal} {Phys. Rev. D}\ }\textbf {\bibinfo {volume}
  {85}},\ \bibinfo {pages} {125028} (\bibinfo {year} {2012}{\natexlab{a}})},\
  \Eprint {https://arxiv.org/abs/1206.0998} {arXiv:1206.0998 [hep-th]}
  \BibitemShut {NoStop}%
\bibitem [{\citenamefont {Manton}\ and\ \citenamefont
  {Sutcliffe}(2004)}]{manton_sutcliffe_2004}%
  \BibitemOpen
  \bibfield  {author} {\bibinfo {author} {\bibfnamefont {N.}~\bibnamefont
  {Manton}}\ and\ \bibinfo {author} {\bibfnamefont {P.}~\bibnamefont
  {Sutcliffe}},\ }\href {https://doi.org/10.1017/CBO9780511617034} {\emph
  {\bibinfo {title} {Topological Solitons}}},\ Cambridge Monographs on
  Mathematical Physics\ (\bibinfo  {publisher} {Cambridge University Press},\
  \bibinfo {year} {2004})\BibitemShut {NoStop}%
\bibitem [{\citenamefont {Weinberg}(2012)}]{weinberg_2012}%
  \BibitemOpen
  \bibfield  {author} {\bibinfo {author} {\bibfnamefont {E.~J.}\ \bibnamefont
  {Weinberg}},\ }\href {https://doi.org/10.1017/CBO9781139017787} {\emph
  {\bibinfo {title} {Classical Solutions in Quantum Field Theory: Solitons and
  Instantons in High Energy Physics}}},\ Cambridge Monographs on Mathematical
  Physics\ (\bibinfo  {publisher} {Cambridge University Press},\ \bibinfo
  {year} {2012})\BibitemShut {NoStop}%
\bibitem [{\citenamefont {Kibble}(1976)}]{Kibble_1976}%
  \BibitemOpen
  \bibfield  {author} {\bibinfo {author} {\bibfnamefont {T.~W.~B.}\
  \bibnamefont {Kibble}},\ }\bibfield  {title} {\bibinfo {title} {Topology of
  cosmic domains and strings},\ }\href
  {https://doi.org/10.1088/0305-4470/9/8/029} {\bibfield  {journal} {\bibinfo
  {journal} {Journal of Physics A: Mathematical and General}\ }\textbf
  {\bibinfo {volume} {9}},\ \bibinfo {pages} {1387} (\bibinfo {year}
  {1976})}\BibitemShut {NoStop}%
\bibitem [{\citenamefont {Abrikosov}(1957)}]{Abrikosov:1956sx}%
  \BibitemOpen
  \bibfield  {author} {\bibinfo {author} {\bibfnamefont {A.~A.}\ \bibnamefont
  {Abrikosov}},\ }\bibfield  {title} {\bibinfo {title} {{On the Magnetic
  properties of superconductors of the second group}},\ }\href@noop {}
  {\bibfield  {journal} {\bibinfo  {journal} {Sov. Phys. JETP}\ }\textbf
  {\bibinfo {volume} {5}},\ \bibinfo {pages} {1174} (\bibinfo {year}
  {1957})}\BibitemShut {NoStop}%
\bibitem [{\citenamefont {Nielsen}\ and\ \citenamefont
  {Olesen}(1973)}]{NIELSEN197345}%
  \BibitemOpen
  \bibfield  {author} {\bibinfo {author} {\bibfnamefont {H.}~\bibnamefont
  {Nielsen}}\ and\ \bibinfo {author} {\bibfnamefont {P.}~\bibnamefont
  {Olesen}},\ }\bibfield  {title} {\bibinfo {title} {Vortex-line models for
  dual strings},\ }\href
  {https://doi.org/https://doi.org/10.1016/0550-3213(73)90350-7} {\bibfield
  {journal} {\bibinfo  {journal} {Nuclear Physics B}\ }\textbf {\bibinfo
  {volume} {61}},\ \bibinfo {pages} {45} (\bibinfo {year} {1973})}\BibitemShut
  {NoStop}%
\bibitem [{\citenamefont {Dunne}(1999)}]{10.1007/3-540-46637-1_3}%
  \BibitemOpen
  \bibfield  {author} {\bibinfo {author} {\bibfnamefont {G.~V.}\ \bibnamefont
  {Dunne}},\ }\bibfield  {title} {\bibinfo {title} {Aspects of chern-simons
  theory},\ }in\ \href@noop {} {\emph {\bibinfo {booktitle} {Aspects
  topologiques de la physique en basse dimension. Topological aspects of low
  dimensional systems}}},\ \bibinfo {editor} {edited by\ \bibinfo {editor}
  {\bibfnamefont {A.}~\bibnamefont {Comtet}}, \bibinfo {editor} {\bibfnamefont
  {T.}~\bibnamefont {Jolic{\oe}ur}}, \bibinfo {editor} {\bibfnamefont
  {S.}~\bibnamefont {Ouvry}},\ and\ \bibinfo {editor} {\bibfnamefont
  {F.}~\bibnamefont {David}}}\ (\bibinfo  {publisher} {Springer Berlin
  Heidelberg},\ \bibinfo {address} {Berlin, Heidelberg},\ \bibinfo {year}
  {1999})\ pp.\ \bibinfo {pages} {177--263}\BibitemShut {NoStop}%
\bibitem [{\citenamefont {Bogomol'nyi}(1976)}]{osti_7309001}%
  \BibitemOpen
  \bibfield  {author} {\bibinfo {author} {\bibfnamefont {E.~B.}\ \bibnamefont
  {Bogomol'nyi}},\ }\bibfield  {title} {\bibinfo {title} {The stability of
  classical solutions},\ }\bibfield  {journal} {\bibinfo  {journal} {Sov. J.
  Nucl. Phys. (Engl. Transl.); (United States)}\ }\textbf {\bibinfo {volume}
  {24:4}},\ \href {https://www.osti.gov/biblio/7309001} {} (\bibinfo {year}
  {1976})\BibitemShut {NoStop}%
\bibitem [{\citenamefont {Paul}\ and\ \citenamefont
  {Khare}(1986)}]{Paul:1986ix}%
  \BibitemOpen
  \bibfield  {author} {\bibinfo {author} {\bibfnamefont {S.~K.}\ \bibnamefont
  {Paul}}\ and\ \bibinfo {author} {\bibfnamefont {A.}~\bibnamefont {Khare}},\
  }\bibfield  {title} {\bibinfo {title} {{Charged Vortices in Abelian Higgs
  Model with Chern-Simons Term}},\ }\href
  {https://doi.org/10.1016/0370-2693(86)91028-2} {\bibfield  {journal}
  {\bibinfo  {journal} {Phys. Lett. B}\ }\textbf {\bibinfo {volume} {174}},\
  \bibinfo {pages} {420} (\bibinfo {year} {1986})},\ \bibinfo {note} {[Erratum:
  Phys.Lett.B 177, 453--453 (1986)]}\BibitemShut {NoStop}%
\bibitem [{\citenamefont {Lee}\ \emph {et~al.}(1990)\citenamefont {Lee},
  \citenamefont {Lee},\ and\ \citenamefont {Min}}]{LEE199079}%
  \BibitemOpen
  \bibfield  {author} {\bibinfo {author} {\bibfnamefont {C.}~\bibnamefont
  {Lee}}, \bibinfo {author} {\bibfnamefont {K.}~\bibnamefont {Lee}},\ and\
  \bibinfo {author} {\bibfnamefont {H.}~\bibnamefont {Min}},\ }\bibfield
  {title} {\bibinfo {title} {Self-dual maxwell chern-simons solitons},\ }\href
  {https://doi.org/https://doi.org/10.1016/0370-2693(90)91084-O} {\bibfield
  {journal} {\bibinfo  {journal} {Physics Letters B}\ }\textbf {\bibinfo
  {volume} {252}},\ \bibinfo {pages} {79} (\bibinfo {year} {1990})}\BibitemShut
  {NoStop}%
\bibitem [{\citenamefont {{Nata Atmaja}}(2017)}]{NATAATMAJA2017351}%
  \BibitemOpen
  \bibfield  {author} {\bibinfo {author} {\bibfnamefont {A.}~\bibnamefont
  {{Nata Atmaja}}},\ }\bibfield  {title} {\bibinfo {title} {A method for bps
  equations of vortices},\ }\href
  {https://doi.org/https://doi.org/10.1016/j.physletb.2017.03.007} {\bibfield
  {journal} {\bibinfo  {journal} {Physics Letters B}\ }\textbf {\bibinfo
  {volume} {768}},\ \bibinfo {pages} {351} (\bibinfo {year}
  {2017})}\BibitemShut {NoStop}%
\bibitem [{\citenamefont {Adam}\ and\ \citenamefont
  {Santamaria}(2016)}]{Adam:2016ipc}%
  \BibitemOpen
  \bibfield  {author} {\bibinfo {author} {\bibfnamefont {C.}~\bibnamefont
  {Adam}}\ and\ \bibinfo {author} {\bibfnamefont {F.}~\bibnamefont
  {Santamaria}},\ }\bibfield  {title} {\bibinfo {title} {{The First-Order
  Euler-Lagrange equations and some of their uses}},\ }\href
  {https://doi.org/10.1007/JHEP12(2016)047} {\bibfield  {journal} {\bibinfo
  {journal} {JHEP}\ }\textbf {\bibinfo {volume} {12}},\ \bibinfo {pages}
  {047}},\ \Eprint {https://arxiv.org/abs/1609.02154} {arXiv:1609.02154
  [hep-th]} \BibitemShut {NoStop}%
\bibitem [{\citenamefont {Atmaja}(2020)}]{Atmaja:2018ddi}%
  \BibitemOpen
  \bibfield  {author} {\bibinfo {author} {\bibfnamefont {A.~N.}\ \bibnamefont
  {Atmaja}},\ }\bibfield  {title} {\bibinfo {title} {{Searching for BPS
  vortices with nonzero stress tensor in the generalized
  Born\textendash{}Infeld\textendash{}Higgs model}},\ }\href
  {https://doi.org/10.1140/epjp/s13360-020-00645-9} {\bibfield  {journal}
  {\bibinfo  {journal} {Eur. Phys. J. Plus}\ }\textbf {\bibinfo {volume}
  {135}},\ \bibinfo {pages} {619} (\bibinfo {year} {2020})},\ \Eprint
  {https://arxiv.org/abs/1807.01483} {arXiv:1807.01483 [hep-th]} \BibitemShut
  {NoStop}%
\bibitem [{\citenamefont {Bazeia}\ \emph
  {et~al.}(2012{\natexlab{b}})\citenamefont {Bazeia}, \citenamefont {Casana},
  \citenamefont {da~Hora},\ and\ \citenamefont {Menezes}}]{PhysRevD.85.125028}%
  \BibitemOpen
  \bibfield  {author} {\bibinfo {author} {\bibfnamefont {D.}~\bibnamefont
  {Bazeia}}, \bibinfo {author} {\bibfnamefont {R.}~\bibnamefont {Casana}},
  \bibinfo {author} {\bibfnamefont {E.}~\bibnamefont {da~Hora}},\ and\ \bibinfo
  {author} {\bibfnamefont {R.}~\bibnamefont {Menezes}},\ }\bibfield  {title}
  {\bibinfo {title} {Generalized self-dual maxwell-chern-simons-higgs model},\
  }\href {https://doi.org/10.1103/PhysRevD.85.125028} {\bibfield  {journal}
  {\bibinfo  {journal} {Phys. Rev. D}\ }\textbf {\bibinfo {volume} {85}},\
  \bibinfo {pages} {125028} (\bibinfo {year} {2012}{\natexlab{b}})}\BibitemShut
  {NoStop}%
\bibitem [{\citenamefont {Dey}(1977)}]{DEY197715}%
  \BibitemOpen
  \bibfield  {author} {\bibinfo {author} {\bibfnamefont {S.}~\bibnamefont
  {Dey}},\ }\bibfield  {title} {\bibinfo {title} {Perturbed iterative solution
  of nonlinear equations with applications to fluid dynamics},\ }\href
  {https://doi.org/https://doi.org/10.1016/0771-050X(77)90020-1} {\bibfield
  {journal} {\bibinfo  {journal} {Journal of Computational and Applied
  Mathematics}\ }\textbf {\bibinfo {volume} {3}},\ \bibinfo {pages} {15}
  (\bibinfo {year} {1977})}\BibitemShut {NoStop}%
\end{thebibliography}%

\end{document}